\documentclass[submission,copyright,creativecommons]{eptcs}
\providecommand{\event}{HCVS 2021} 
\usepackage{breakurl}             
\usepackage{underscore}           

\usepackage[spanish,english]{babel}
\usepackage{xspace}
\usepackage{amssymb}
\usepackage{amsmath}
\usepackage{tikz}
\usetikzlibrary{decorations,arrows,arrows.meta,shapes,patterns}
\usepackage{paralist}
\usepackage{listings}

\let\citeN\cite

\newcommand{\irank}[0]{i\textsc{RankFinder}\xspace}
\newcommand{\smt}[0]{\textsc{SMT}\xspace}
\newcommand{\chc}[0]{{CHC}\xspace}
\newcommand{\chcs}[0]{CHCs\xspace}
\newcommand{\ts}[0]{TS\xspace}
\newcommand{\tss}[0]{TSs\xspace}

\newcommand{\cfgs}[0]{CFGs\xspace}
\newcommand{\slc}[0]{SLC\xspace}

\newcommand{\node}[1]{\ensuremath{\mathtt{#1}}}

\newcommand{\trsys}[0]{\ensuremath{{\cal T}}\xspace}
\newcommand{\cbox}[1]{\colorbox{yellow}{\textcolor{blue}{#1}}}
\newcommand{\lrf}[0]{LRF\xspace}
\newcommand{\lrfs}[0]{LRFs\xspace}
\newcommand{\llrf}[0]{LLRF\xspace}
\newcommand{\llrfs}[0]{LLRFs\xspace}

\newcommand{\mlrf}[0]{M{$\Phi$}RF\xspace}
\newcommand{\mlrfs}[0]{M{$\Phi$}RFs\xspace}
\newcommand{\cfr}[0]{CFR\xspace}

\let\vect=\vec
\renewcommand{\vec}[1]{\mathbf{#1}}
\newcommand{\tuple}[1]{\langle #1 \rangle}
\newcommand{\rats}{\ensuremath{\mathbb Q}\xspace}
\newcommand{\transitions}{\poly{Q}}
\newcommand{\stransitions}{\scriptscriptstyle\poly{Q}}
\newcommand{\trcv}[2]{\ensuremath{\bigl(\begin{smallmatrix}{#1}\hfill\\{#2}\hfill\end{smallmatrix}\bigr)}}
\newcommand{\poly}[1]{{\mathcal #1}}
\newcommand{\ptime}[0]{\ensuremath{\mathtt{PTIME}}\xspace}
\newcommand{\coNP}[0]{\ensuremath{\mathtt{coNP}}\xspace}

\lstdefinestyle{general}{ 
  stringstyle=\ttfamily,
  showstringspaces = false,
  basicstyle=\linespread{0.9}\ttfamily,
  commentstyle=\small\emph,
  keywordstyle=\small\bfseries,
  numbers=left,
  mathescape=true,
  numbersep=3pt,
  numberstyle=\tiny,
  numberfirstline=true,
  breaklines=true,
  language=C,
  xleftmargin=1em,
  morekeywords={if,then,else,while,do,return,length,int,void},
  columns=[l]flexible,
  tabsize=4
}

\title{Termination Analysis of Programs with Multiphase Control-Flow%
  \thanks{%
    This paper is a summary of the corresponding invited talk of the
    second author given at HCVS 2021, and is based on the PhD thesis
    of the first author~\cite{Domenech21} and on collaborations with
    Amir M. Ben-Amram~\cite{Ben-AmramG17,Ben-AmramDG19} and John
    P. Gallagher~\cite{DomenechGG19}.
This work was funded partially by the Spanish MCIU, AEI and FEDER (EU)
project RTI2018-094403-B-C31, by the CM project S2018/TCS-4314
co-funded by EIE Funds of the European Union, and by the UCM
CT42/18-CT43/18 grant. }}

\author{
  Jes\'us J. Domenech$^1$ \quad\quad\quad Samir Genaim$^{1,2}$\\
  \institute{$^1$ Universidad Complutense de Madrid, Spain}%
  \institute{$^2$ Instituto de Tecnolog\'ia del Conocimiento, Madrid, Spain}%
  \email{jdomenech@ucm.es \quad\quad\quad sgenaim@ucm.es}%
}

\def\titlerunning{Termination
  Analysis of Programs with Multiphase Control-Flow}
\def\authorrunning{Jes\'us J. Domenech and Samir Genaim}

\begin{document}
\maketitle

\begin{abstract}
Programs with multiphase control-flow are programs where the execution
passes through several~(possibly implicit) phases. Proving termination
of such programs (or inferring corresponding runtime bounds) is often
challenging since it requires reasoning on these phases separately.
In this paper we discuss techniques for proving termination of such
programs, in particular: (1) using multiphase ranking functions, where
we will discuss theoretical aspects of such ranking functions for
several kinds of program representations; and (2) using control-flow
refinement, in particular partial evaluation of Constrained Horn
Clauses, to simplify the control-flow allowing, among other things, to
prove termination with simpler ranking functions.
\end{abstract}

\section{Introduction}
\label{sec:intro}

Proving that a program will eventually terminate, i.e., that it does
not go into an infinite loop, is one of the most fundamental tasks of
program verification, and has been the subject of voluminous research.
Perhaps the best known, and often used, technique for proving
termination is that of \emph{ranking functions}, which has already
been used by Alan Turing in his early work on program
verification~\cite{Turing:1949:CLR}.
This consists of finding a function $\rho$ that maps program states
into the elements of a well-founded ordered set, such that
$\rho(s) \succ \rho(s')$ holds for any consecutive states $s$ and
$s'$.
This implies termination since infinite descent is impossible in a
well-founded order.
Besides proving termination, Turing~\citeN{Turing:1949:CLR} mentions
that ranking functions can be used to bound the length computations as
well.
This is useful in applications such as cost analysis and loop
optimisation~\cite{Feautrier92.1,ADFG:2010,AlbertAGP11,BrockschmidtE0F16}.

Unlike termination of programs in general, which is undecidable, the
algorithmic problems of detection (deciding the existence) or
generation (synthesis) of a ranking function can well be solvable,
given certain choices of the program representation, and the class of
ranking functions.
There is a considerable amount of research in this direction, in which
different kinds of ranking functions for different kinds of program
representations were considered.
In some cases, the algorithmic problems have been completely settled,
and efficient algorithms provided, while other cases remain open.

A common program representation in this context is \emph{Single-path
  Linear-Constraint} (\slc) loops, where a state is described by the
values of numerical variables, and the effect of a transition~(one
iteration) is described by a conjunction of \emph{linear constraints}.
Here is an example of this loop representation; primed variables
$x'$ and $y'$ refer to the state following the transition:
\begin{equation}
\label{eq:intro:loop}
while~(x \le y)~do~x' = x+1,\, y' \le y
\end{equation}
Note that $x'=x+1$ is an equation, not an assignment.  The description
of a loop may involve linear inequalities rather than equations, such
as $y'\le y$ above, and consequently, be non-deterministic.
Note that for a \slc loop with $n$ variables, a transition can be seen
as a point $\trcv{\vec{x}}{\vec{x}'} \in \rats^{2n}$, where its first
$n$ components correspond to $\vec{x}$ and its last $n$ components to
$\vec{x}'$. We denote the set of all transitions by $\transitions$,
which is a polyhedron.

A more general program representation is \emph{Transition Systems}
(\tss), which are defined by \emph{Control-Flow Graphs} (\cfgs) with
numerical variables, consisting of nodes representing program
locations and edges annotated with linear constraints (polyhedra)
describing how values of variables change when moving from one
location to another.
Figure~\ref{fig:phases1} includes a program and its corresponding \ts
$\trsys$ to its right. Primed variables in the linear constraints
refer to the state following the transition, exactly as in the case of
\slc loops.

In both program representations mentioned above, the domain of
variables is also important as it typically affects the complexity of
the underlying decision and synthesis problems.
Although these program representations allow only numerical variables
and linear constraints, data structures can be handled using
\emph{size abstractions}, e.g., length of lists, depth of trees,
etc.~\cite{LS:97,DBLP:conf/popl/LeeJB01,DBLP:journals/toplas/BruynoogheCGGV07,DBLP:journals/toplas/SpotoMP10,DBLP:conf/popl/MagillTLT10,aprove17}. In
such case, variables represent sizes of corresponding data structures.

\begin{figure}[t]
\begin{center}
\begin{minipage}{0.48\textwidth}
\begin{lstlisting}
void phases1(int x, int y, int z) {
  while( x >= 1 ){
    if ( y <= z - 1 ) {
      y = y + 1;
    } else {
      x = x - 1;
    }
  }
}
\end{lstlisting}
\end{minipage}
\begin{minipage}{0.2\textwidth}
  \fbox{
    \scalebox{.78}{

\begin{tikzpicture}[>=latex,line join=bevel,]
  \node (n0) at (12.5bp,131.0bp) [draw,fill=green,circle] {\scalebox{1.2}{$\node{n_0}$}};
  \node (n1) at (12.5bp,70.0bp) [draw,circle] {\scalebox{1.2}{$\node{n_1}$}};
  \node (n2) at (78.5bp,70.0bp) [draw,circle] {\scalebox{1.2}{$\node{n_2}$}};
  \node (n3) at (12.5bp,9.0bp) [draw,circle] {\scalebox{1.2}{$\node{n_3}$}};
  \node (trs) at (78.5bp,9.0bp) [] {\scalebox{1.2}{\cbox{$\trsys$}}};
  \path (n0) edge [->] node[left] {\scalebox{1.2}{$\stransitions_0$}} (n1);
  \path (n1) edge [->] node[above] {\scalebox{1.2}{$\stransitions_1$}} (n2);
  \path (n1) edge [->] node[left] {\scalebox{1.2}{$\stransitions_2$}} (n3);
  \draw [->] (n2) ..controls (68.895bp,57.1bp) and (63.379bp,51.598bp)  .. (57.0bp,49.0bp) .. controls (47.533bp,45.144bp) and (43.467bp,45.144bp)  .. (34.0bp,49.0bp) .. controls (30.91bp,50.259bp) and (28.023bp,51.198bp)  .. (n1);
  \draw (45.5bp,58.0bp) node {\scalebox{1.2}{$\stransitions_3$}};
  \draw [->] (n2) ..controls (68.895bp,82.9bp) and (63.379bp,88.402bp)  .. (57.0bp,91.0bp) .. controls (47.533bp,94.86bp) and (43.467bp,94.86bp)  .. (34.0bp,91.0bp) .. controls (30.91bp,89.74bp) and (28.023bp,87.802bp)  .. (n1);
  \draw (45.5bp,106.0bp) node {\scalebox{1.2}{$\stransitions_4$}};
\end{tikzpicture}
 }
  }
\end{minipage}
\begin{minipage}{0.3\textwidth}
  \fbox{
    \scalebox{.65}{

\begin{tikzpicture}[>=latex,line join=bevel,]
  \node (n0) at (14.5bp,159.89bp) [draw,fill=green,circle] {\scalebox{1.3}{$\node{n_0}$}};
  \node (n13) at (78.5bp,159.89bp) [draw,circle] {\scalebox{1.3}{$\node{n_1^3}$}};
  \node (n32) at (144.5bp,159.89bp) [draw,circle] {\scalebox{1.3}{$\node{n_3^2}$}};

  \node (n22) at (78.5bp,91.89bp) [draw,circle] {\scalebox{1.3}{$\node{n_2^2}$}};
  \node (n12) at (12.5bp,91.89bp) [draw,circle] {\scalebox{1.3}{$\node{n_1^2}$}};
  \node (trs) at (144.5bp,91.89bp) [] {\scalebox{1.3}{\cbox{$\trsys_{pe}$}}};

  \node (n21) at (12.5bp,23.892bp) [draw,circle] {\scalebox{1.3}{$\node{n_2^1}$}};
  \node (n31) at (144.5bp,23.892bp) [draw,circle] {\scalebox{1.3}{$\node{n_3^1}$}};
  \node (n11) at (78.5bp,23.892bp) [draw,circle] {\scalebox{1.3}{$\node{n_1^1}$}};

  \path (n0) edge [->] node[above] {\scalebox{1.3}{$\stransitions_0$}} (n13);
  \path (n13) edge [->] node[right] {\scalebox{1.3}{$\stransitions_1$}} (n22);
  \path (n13) edge [->] node[above] {\scalebox{1.3}{$\stransitions_2$}} (n32);
  \path (n22) edge [->] node[right] {\scalebox{1.3}{$\stransitions_6$}} (n11);
  \path (n22) edge [->] node[above] {\scalebox{1.3}{$\stransitions_5$}} (n12);
  \path (n11) edge [->] node[above] {\scalebox{1.3}{$\stransitions_7$}} (n21);
  \path (n11) edge [->] node[above] {\scalebox{1.3}{$\stransitions_8$}} (n31);

  \draw [->] (n21) ..controls (23.796bp,9.3918bp) and (28.588bp,5.0967bp)  .. (34.0bp,2.8922bp) .. controls (43.467bp,-0.96406bp) and (47.533bp,-0.96406bp)  .. (57.0bp,2.8922bp) .. controls (59.199bp,3.7878bp) and (61.295bp,5.0284bp)  .. (n11);
  \draw (45.5bp,11.892bp) node {\scalebox{1.3}{$\stransitions_{6}$}};

  \draw [->] (n12) ..controls (23.796bp,77.392bp) and (28.588bp,73.097bp)  .. (34.0bp,70.892bp) .. controls (43.467bp,67.036bp) and (47.533bp,67.036bp)  .. (57.0bp,70.892bp) .. controls (59.199bp,71.788bp) and (61.295bp,73.028bp)  .. (n22);
  \draw (45.5bp,79.892bp) node {\scalebox{1.3}{$\stransitions_{1}$}};

\end{tikzpicture}
 }
  }
  \end{minipage}
\end{center}

\bigskip
\begin{minipage}{\textwidth}
\[
\begin{array}{r@{}llllll}
\transitions_0\equiv&  \{&&x'=x,&y'=y,&z'=z&\} \\
\transitions_1\equiv&  \{x\ge 1,&& x'=x,&y'=y,&z'=z&\} \\
\transitions_2\equiv&  \{x\leq 0,&& x'=x,&y'=y,&z'=z&\} \\
\transitions_3\equiv&  \{y\leq z-1,&& x'=x,& y'=y+1,& z'=z&\} \\
\transitions_4\equiv&  \{y\geq z,&& x'=x-1,& y'=y,& z'=z&\} \\
\transitions_5\equiv&  \{ x\ge 1,& y\leq z-1,& x' = x,& y'=y+1,& z'=z&\} \\
\transitions_6\equiv&  \{x\ge 1,& y\geq z,& x'=x-1,& y'=y,& z'=z&\} \\
\transitions_7\equiv&  \{x\ge 1,& y\geq z,& x'=x,&y'=y,&z'=z&\}  \\
\transitions_8\equiv&  \{x\leq 0,& y\geq z,& x'=x,&y'=y,&z'=z&\} \\
\end{array}
\]
\end{minipage}

\caption{A loop with $2$ phases, it corresponding \ts $\trsys$, and the \ts $\trsys_{pe}$ after applying \cfr.}
\label{fig:phases1}

\end{figure}

Due to practical considerations, termination analysis tools typically
focus on classes of ranking functions that can be synthesised
efficiently.
This, however, does not mean that theoretical aspects of such classes
are put aside, as understanding theoretical limits and properties of
the underlying problems is necessary for developing practical
algorithms.
The most popular class of ranking functions in this context is
probably that of \emph{Linear Ranking Functions} (\lrfs).
A \lrf is a function
$\rho(x_1,\dots,x_n) = a_1x_1+\dots+a_n x_n + a_0$ such that any
transition from $\vec{x}$ to $\vec{x}'$ satisfies
\begin{inparaenum}[(i)]
\item\label{intro:lrf1} $\rho(\vec{x}) \ge 0$; and
\item\label{intro:lrf2} $\rho(\vec{x})-\rho(\vec{x}') \ge 1$.
\end{inparaenum}
For example, $\rho(x,y)=y-x$ is a \lrf for 
Loop~\eqref{eq:intro:loop}.
Several polynomial-time algorithms to find a \lrf using linear
programming
exist~\cite{ADFG:2010,DBLP:conf/tacas/ColonS01,Feautrier92.1,DBLP:journals/tplp/MesnardS08,DBLP:conf/vmcai/PodelskiR04,DBLP:conf/pods/SohnG91}. These
algorithms are complete%
\footnote{Complete means that if there is a \lrf, they will find one.}
for \tss with rational-valued variables, but not with integer-valued
variables. Ben-Amram and Genaim~\citeN{Ben-AmramG13popl} showed how
completeness for the integer case can be achieved, and also classified
the corresponding decision problem as co-NP complete.

Despite their popularity, \lrfs do not suffice for all programs, and a
natural question is what to do when a \lrf does not exist; and a
natural answer is to try a richer class of ranking functions. Of
particular importance is the class of \emph{Lexicographic-Linear
  Ranking Functions} (\llrfs). These are tuples of linear functions
that decrease lexicographically over a corresponding well-founded
ordered set.
For example, the program depicted in Figure~\ref{fig:phases1} does
not have a \lrf, but it has the \llrf $\tuple{z-y,x}$ since: for the
\emph{then} branch $z-y$ decreases and for the \emph{else} branch $x$
decreases while $z-y$ does not change.
\llrfs might be necessary even for \slc loops. For example, the
following \slc loop
\begin{equation}
\label{eq:intro:loop_xyz}
while~(x \ge 0)~do~x' = x+y,\, y' = y+z, z'=z-1
\end{equation}
does not have a \lrf, but can be proved terminating using the \llrf
$\tuple{z,y,x}$.

There are several definitions for \llrfs in the
literature~\cite{ADFG:2010,Ben-AmramG13jv,DBLP:conf/cav/BradleyMS05,LarrazORR13}
and they have different power, i.e., some can prove termination of a
program while others fail.
They have corresponding polynomial-time synthesis algorithms for the
case of rational variables, and the underlying decision problems are
co-NP complete for the case of integer
variables~\cite{Ben-AmramG13jv,Ben-AmramG15}.
The definition of Larraz et al.~\citeN{LarrazORR13} is the most
general in this spectrum of \llrfs, its complexity classification is
not known yet and corresponding complete synthesis algorithms do not
exist.

The \llrf $\tuple{z,y,x}$ used above for
Loop~\eqref{eq:intro:loop_xyz} belongs to a class of \llrfs that is
know as \emph{Multiphase-Linear Ranking Functions} (\mlrfs).
Ranking functions in this class are characterised by the following
behaviour: the first component \emph{always} decreases, and when it
becomes negative the second component starts to decrease (and will
keep decreasing during the rest of the execution), and when it becomes
negative the third component starts to decrease, and so on. This
behaviour defines phases through which executions pass.
This is different from \llrfs in general since once a component
becomes negative it cannot be used anymore.
Also the \llrf $\tuple{z-y,x}$ for the program of
Figure~\ref{fig:phases1} induces a similar multiphase behaviour, since
once $z-y$ becomes negative it cannot be used anymore. However, it is
not a \mlrf since $z-y$ stops decreasing once we move to the second
phase.

In the rest of this paper we overview works on termination analysis of
multiphase programs, in particular: in Section~\ref{sec:mlrfs} we
discuss techniques for proving termination using \mlrfs; and in
Section~\ref{sec:cfr} we discuss techniques for proving termination
using \emph{Control-Flow Refinement} (\cfr). Section~\ref{sec:conc}
ends the paper with some concluding remarks and discusses further
research directions.

\section{Termination analysis using multiphase ranking functions}
\label{sec:mlrfs}

A \mlrf is a tuple $\tuple{\rho_1,\ldots,\rho_d}$ of linear functions
that define phases of the program that are linearly ranked, where $d$
is the \emph{depth} of the \mlrf, intuitively the number of phases.
The decision problem \emph{Existence of a \mlrf} asks to determine
whether a program has a \mlrf. The \emph{bounded} decision problem
restricts the search to \mlrfs of depth $d$, where $d$ is part of the
input.

For the case of \slc loops, the complexity and algorithmic aspects of
the bounded version of the \mlrf problem were settled by Ben-Amram and
Genaim~\citeN{Ben-AmramG17}.
The decision problem is \ptime for \slc loops with rational-valued
variables, and \coNP-complete for \slc loops with integer-valued
variables; synthesising \mlrfs, when they exist, can be performed in
polynomial and exponential time, respectively.
We note that the proof of the rational case is done by showing that
\mlrfs and \emph{nested} ranking functions~\cite{LeikeHeizmann15} (a
strict subclass of \mlrfs for which a polynomial-time algorithm exists)
have the same power for \slc loops.
Besides, they show that for \slc loops \mlrfs have the same power as
general lexicographic-linear ranking functions, and that \mlrfs induce
linear iteration bounds.
The problem of deciding if a given \slc loop admits a \mlrf, without a
given bound on the depth, is still open.

In practice, termination analysis tools search for \mlrfs
incrementally, starting by depth $1$ and increase the depth until they
find one, or reach a predefined limit, after which the returned answer
is \emph{don't know}.
Finding a theoretical upper-bound on the depth of a \mlrf, given the
loop, would also settle this problem, however, as shown by Ben-Amram and
Genaim~\citeN{Ben-AmramG17} such bound must depend not only on the
number of constraints or variables, as for other classes of
\llrfs~\cite{ADFG:2010,Ben-AmramG13jv,DBLP:conf/cav/BradleyMS05}, but
also on the coefficients used in the corresponding constraints.
Yuan et al.~\cite{YuanLS21} proposed an incomplete method to bound the
depth of \mlrfs for \slc loops.

Ben-Amram et al.~\cite{Ben-AmramDG19} have done a significant progress
towards solving the problem of \emph{existence of a \mlrf} for \slc
loop, i.e., seeking a \mlrf without a given bound on the depth. In
particular, they present an algorithm for seeking \mlrfs that reveals
novel insights on the structure of these ranking functions.
In a nutshell, the algorithm starts from the set of transitions of the
given \slc loop, which is a polyhedron, and iteratively removes
transitions $\trcv{\vec{x}}{\vec{x}'}$ such that
$\rho(\vec{x})-\rho(\vec{x}')>0$ for some function
$\rho(\vec{x})=\vect{a}\cdot\vec{x}+b$ that is \emph{non-negative on
  all enabled states}.
The process continues iteratively, since after removing some
transitions, more functions $\rho$ may satisfy the non-negativity
condition, and they may eliminate additional transitions in the next
iteration.
When all transitions are eliminated in a finite number of iterations,
one can construct a \mlrf using the $\rho$ functions; and when
reaching a situation in which no transition can be eliminated, the
remaining set of transitions, which is a polyhedron, is actually a
recurrent set that witnesses non-termination.
The algorithm always finds a \mlrf if one exists, and in many cases, it
finds a recurrent set when the loop is non-terminating, however, it is
not a decision procedure as it diverges in some cases.
Nonetheless, the algorithm provides important insights into the
structure of \mlrfs. Apart from revealing a relation between seeking
\mlrfs and seeking recurrent sets, these insights are useful for
finding classes of \slc loops for which, when terminating, there is
always a \mlrf and thus have linear run-time bound.
This result was proven for two kinds of \slc loops, both considered in
previous work, namely \emph{octagonal relations} and \emph{affine
  relations with the finite-monoid property} -- for both classes,
termination has been proven decidable~\cite{BIKtacas2012jv}.

Ben-Amram et al.~\cite{Ben-AmramDG19} have also suggested a new
representation for \slc loops, called the \emph{displacement}
representation, that provides new tools for studying termination of
\slc loops in general, and the existence of a \mlrf in particular.
In this representation, a transition $\trcv{\vec{x}}{\vec{x}'}$ is
represented as $\trcv{\vec{x}}{\vec{y}}$ where
$\vec{y}=\vec{x}'-\vec{x}$.
Using this representation the algorithm described above can be
formalised in a simple way that avoids computing the $\rho$ functions
mentioned above, and reduces the existence of a \mlrf of depth $d$ to
unsatisfiability of a certain linear constraint system.  Moreover, any
satisfying assignment for this linear constraint system is a witness
that explains why the loop has no \mlrf of depth $d$.
As an evidence on the usefulness of this representation in general,
they showed that some non-trivial observations on termination of
bounded \slc loops (i.e., the set of transitions is a bounded
polyhedron) are made straightforward in this representation, while
they are not easy to see in the normal representation, in particular:
a bounded \slc loop terminates if and only if it has a \lrf, and it
does not terminate if and only if it has a fixpoint transition
$\trcv{\vec{x}}{\vec{x}}$.

The works discussed above are limited to the case of \slc loops. The
case of general \tss has been considered by Leike and
Heizmann~\citeN{LeikeHeizmann15} and Li et al.~\citeN{li2016depth},
where both translate the existence of a \mlrf of a given depth $d$ to
solving a corresponding non-linear constraint problem, which is
complete for the case of rational variables. However, while complete,
these approaches do not provide any insights on the complexity of the
underlying decision problems.
The technique of Borralleras et al.~\citeN{BorrallerasBLOR17} can be
used to infer, among other things, \mlrfs for \tss without a given
bound on the depth. It is based on solving corresponding safety
problems using Max-\smt, and it is not complete.

\section{Termination analysis using control-flow refinement}
\label{sec:cfr}

As we have seen in Section~\ref{sec:intro}, there are \tss with
multiphase behaviour that do not admit \mlrfs, but rather a different
notion of \llrf that does not require the components to keep
decreasing after turning negative.
To prove termination of such \tss one can use other classes of
\llrfs~\cite{ADFG:2010,Ben-AmramG13jv,DBLP:conf/cav/BradleyMS05,LarrazORR13},
however, this might not be enough due to complex control-flow where
abstract properties of several implicit execution paths are merged.
To overcome this imprecision, not only for termination but for program
analysis in general, one approach is to simplify the control-flow in
order to make implicit execution paths explicit, and thus makes it
possible to infer the desired properties with a weaker version of the
analysis. This is known as \emph{Control-Flow Refinement} (\cfr).

Let us see how \cfr can simplify the kind of ranking functions needed
to proved termination, and thus improve the precision of the
underlying termination analyser.
Consider the program depicted Figure~\ref{fig:phases1} again, and
recall that we failed to prove its termination when using only \lrfs.
Examining this program carefully, we can see that any execution passes
in two phases: in the first one, $\mathtt{y}$ is incremented until it
reaches the value of $\mathtt{z}$, and in the second phase
$\mathtt{x}$ is decremented until it reaches $\mathtt{0}$.
Lets us transform the program into a semantically equivalent one such
that the two phases are separate and explicit:
\begin{lstlisting}
  while (x >= 0 and y <= z-1) y = y + 1;
  while (x >= 0 and y >= z) x = x - 1;
\end{lstlisting}

\noindent
Now we can prove termination of this program using \lrfs only: for the
first loop $z-y$ is a \lrf, and for the second loop $x$ is a
\lrf. Moreover, cost analysis tools that are based on bounding loop
iterations using \lrfs~\cite{AlbertAGP11} would infer a linear bound
for this program while they would fail on the original one.

Apart from simplifying the termination proof, there are also cases
where it is not possible to prove termination without such
transformations even when using \llrfs. In addition, \cfr can help in
inferring more precise invariants, without the need for expensive
disjunctive abstract domains, which can benefit any analysis that uses
such invariants, e.g., termination and cost analysis.

\cfr has been considered by Gulwani et al.~\citeN{GulwaniJK09} and
Flores{-}Montoya and H{\"{a}}hnle~\citeN{Flores-MontoyaH14} to improve
the precision of cost analysis, and by Sharma et
al.~\citeN{SharmaDDA11} to improve the precision of invariants in
order to prove program assertions. While all these techniques can
automatically obtain the transformed program above, they are developed
from scratch and tailored to some analysis of interest.
Recently, \cfr has also been considered by Albert et
al.~\citeN{AlbertBBMR19} to improve cost analysis as well, but from a
different perspective that uses termination witnesses to guide \cfr.

Since \cfr is, in principle, a program transformation that specialises
programs to distinguish different execution scenarios, Domenech et
al.~\citeN{DomenechGG19} explored the use of general-purpose
specialisation techniques for \cfr, in particular the techniques of
Gallagher~\citeN{Gallagher19} for partial evaluation of Constrained
Horn Clauses~(\chcs).
Basing \cfr on partial evaluation has the clear advantage that
soundness comes for free because partial evaluation guarantees
semantic equivalence between the original program and its transformed
version.
Moreover, this way we obtain a \cfr procedure that is not tailored for
a particular purpose, but rather can be tuned depending on the
application domain.

Domenech et al.~\cite{DomenechGG19} developed such a \cfr procedure
for \tss, by transforming \tss into \chcs and using the partial
evaluator of Gallagher~\citeN{Gallagher19}, and integrated it in a
termination analysis algorithm in a way that allows applying \cfr at
different levels of granularity, and thus controlling the trade-off
between precision and performance.
This is done by suggesting different schemes for applying \cfr, not
only as a preprocessing step but also on specific parts of the \ts
which we could not prove terminating.
Moreover, they developed heuristics for automatically configuring
partial evaluation (i.e., inferring properties to guides
specialisation) in order to achieve the desired \cfr.
Experimental evaluation provides a clear evidence to that their \cfr
procedure significantly improves the precision of termination
analysis, cost analysis, and invariants generation.

Let use demonstrate this \cfr procedure on the \ts $\trsys$ that is
depicted in Figure~\ref{fig:phases1}.
In a first step, the \ts $\trsys$ is translated into the following
(semantically) equivalent \chc program:
\[
\begin{array}{rlllll}
q_{\node{n_0}}(x,y,z) \leftarrow& q_{\node{n_1}}(x,y,z). \\
q_{\node{n_1}}(x,y,z) \leftarrow& \{ x\ge 1\},& q_{\node{n_2}}(x,y,z). \\
q_{\node{n_1}}(x,y,z) \leftarrow& \{ x\le 0\},& q_{\node{n_3}}(x,y,z). \\ 
q_{\node{n_2}}(x,y,z) \leftarrow& \{ y\le z-1,&y'=y+1\},& q_{\node{n_1}}(x,y',z). \\
q_{\node{n_2}}(x,y,z) \leftarrow& \{ y\ge z,&x'=x-1\},& q_{\node{n_1}}(x',y,z). \\
\end{array}
\]
Then the partial evaluator of Gallagher~\citeN{Gallagher19} is
applied, using the (automatically inferred) properties
$\{x\ge 1, y\ge z\}$ for the loop head predicate
$q_{\node{n_1}}(x,y,z)$, which results in the following \chc program:
\[
\begin{array}{rllllll}
q_{\node{n_0}}(x,y,z)   \leftarrow&  q_{\node{n_1^3}}(x,y,z).\\
q_{\node{n_1^3}}(x,y,z) \leftarrow&\{  x\le 0\},& q_{\node{n_3^2}}(x,y,z).\\
q_{\node{n_1^3}}(x,y,z) \leftarrow&\{  x\ge 1\},& q_{\node{n_2^2}}(x,y,z).\\
q_{\node{n_2^2}}(x,y,z) \leftarrow&\{  x\ge 1,& y\le z-1,&y'=y+1\},& q_{\node{n_1^2}}(x,y',z).\\
q_{\node{n_2^2}}(x,y,z) \leftarrow&\{  x\ge 1,& y\ge z,&x'=x-1\},& q_{\node{n_1^1}}(x',y,z).\\
q_{\node{n_1^2}}(x,y,z) \leftarrow&\{  x\ge 1\},& q_{\node{n_2^2}}(x,y,z).\\
q_{\node{n_1^1}}(x,y,z) \leftarrow&\{  x\le 0,& y\ge z\},& q_{\node{n_3^1}}(x,y,z).\\
q_{\node{n_1^1}}(x,y,z) \leftarrow&\{  x\ge 1,& y\ge z\},& q_{\node{n_2^1}}(x,y,z).\\
q_{\node{n_2^1}}(x,y,z) \leftarrow&\{  x\ge 1,& y\ge z,&x'=x-1\},& q_{\node{n_1^1}}(x',y,z).\\
\end{array}
\]
Now translating this \chc program into a \ts results in the \ts
$\trsys_{pe}$ that is depicted in Figure~\ref{fig:phases1}. Note that
the two phases are now separated into two different strongly connected
components, which can be proven terminating using only \lrfs.

\section{Concluding Remarks}
\label{sec:conc}

In this paper we have discussed techniques for proving termination of
programs with multiphase control-flow, where the execution passes
through several~(possibly implicit) phases.
In particular, we have discussed techniques that correspond to our
recent work: (1) the use of multiphase ranking
functions~\cite{Ben-AmramG17,Ben-AmramDG19}; and (2) the use of
control-flow refinement~\cite{DomenechGG19}.
As a byproduct of our research, we have developed an open-source
termination analyser called
\irank\footnote{\url{http://irankfinder.loopkiller.com}} that
implements all these techniques as well as other state-of-the-art
techniques. Some of the components of \irank can be used
independently, in particular the \cfr component that can be used to
incorporate \cfr in static analysers with little effort.

\paragraph{Future Work.}
For \mlrfs, an obvious future direction is to study the problem of
deciding whether a \ts has a \mlrf, both from algorithmic and
theoretical complexity perspectives.
In initial unpublished work, we have proven that the corresponding
decision problem is NP-hard for the rational setting, but we could not
obtain a further classification.
Further exploration of the \mlrf problem for \slc loops is also
required since it is not solved for the general case yet.
For \cfr, we have developed some heuristics for the automatic
inference of properties, which is crucial for obtaining the desired
transformations, but further research in this direction is required.
We concentrated on \cfr of \tss, and in a future direction one could
apply our \cfr techniques for program representations that allow
recursion as well. Technically, this would not require much work since
the partial evaluation techniques of Gallagher~\citeN{Gallagher19}
specialise \chcs that include recursion already.
We also concentrated on numerical programs, and a possible future
direction can concentrate on using \cfr for program analysis tools
where the data is not necessarily numerical. Here one should also
adapt the partial evaluation techniques to support such
specialisations, which seems doable for the partial evaluation
techniques of Gallagher~\citeN{Gallagher19} since it is based on using
abstract properties like those used in abstract domains of program
analysis.

\bibliographystyle{eptcs}

\end{document}